\documentclass[12pt]{article}
\usepackage{graphicx}
\usepackage{amssymb}
\usepackage{amsmath}

\begin{document}

\begin{titlepage}

\begin{flushright}
ICRR-Report 636-2012-25\\
IPMU 12-0211\\
TU-925
\end{flushright}

\vskip 1.35cm

\begin{center}

{\large 
{\bf Hubble-induced mass from MSSM plasma} \\
}

\vskip 1.2cm

Masahiro Kawasaki$^{a,c}$,
Fuminobu Takahashi$^{b,c}$
and Tomohiro Takesako$^a$ \\

\vskip 0.4cm

{ \it$^a$Institute for Cosmic Ray Research,
University of Tokyo, Kashiwa 277-8582, Japan}\\
{\it $^b$Department of Physics, Tohoku University, Sendai 980-8578, Japan}\\
{\it $^c$Kavli Institute for the Physics and Mathematics of the Universe,
University of Tokyo, Kashiwa 277-8568, Japan}\\
\date{\today}

\begin{abstract} 
We evaluate the effective mass of a scalar field $\phi$ coupled to thermal plasma through Planck-suppressed interactions.
We find it useful to rescale the coupled fields so that all the $\phi$-dependences are absorbed into the yukawa and gauge couplings,
which allows us to read off the leading order contributions to the effective mass $\tilde m_{\phi}$ from the 2-loop free energy calculated
with the rescaled couplings. 
We give an analytical expression for $\tilde m_{\phi}$ at a sufficiently high temperature
in the case where $\phi$ is coupled to the MSSM chiral superfields through non-minimal 
K$\ddot{\text{a}}$hler potential.
We find that $|\tilde m_{\phi}^2|$ is about $10^{-3} H^2 \sim 10^{-2} H^2$ at the leading order in terms of the couplings for typical parameter sets, where $H$ is the Hubble expansion rate in the radiation-dominated era.
\end{abstract}

\end{center}
\end{titlepage}

\section{Introduction}
Supersymmetry (SUSY) is an attractive candidate  for the physics  beyond the standard model (SM).
Its local version, supergravity,  leads to various  phenomena in cosmology.
In particular, when the inflaton dominates the Universe during and after inflation, 
the supergravity effect  induces an effective mass of order $H$,  the Hubble expansion 
rate, for a general scalar field $\phi$ coupled to the inflaton sector by Planck-suppressed interactions~\cite{Ovrut:1983my, Dine:1983ys, Copeland:1994vg}, 
unless its mass is protected by some symmetry.  Such an effective mass of order  $H$ is called a {\it Hubble-induced mass} and plays an important role in many cosmological scenarios.
For example, a negative Hubble-induced mass enables the Affleck-Dine baryogenesis mechanism~\cite{Affleck:1984fy, Dine:1995uk}.
The enhanced Hubble-induced mass also is  a key for solving the cosmological moduli problem~\cite{Linde:1996cx,Nakayama:2011wqa}.
On the other hand, the Hubble-induced mass will be a main obstacle 
for implementing the curvaton mechanism~\cite{Lyth:2001nq} in supergravity and it 
must be suppressed at least by about one order of magnitude.  

There have been arguments about whether or not such an effective mass of order $H$ arises during 
 the radiation-dominated (RD) era after reheating~\cite{hep-ph/0402174, Kawasaki:2011zi}.
In Ref.~\cite{hep-ph/0402174}, it was claimed that the effective mass for $\phi$, which is coupled to thermal plasma through Planck-suppressed
interactions,  is suppressed by the ratio of the zero-temperature mass of the particles in the plasma and the plasma temperature, and therefore it 
is much smaller than $H$. However, the dispersion relation of the thermalized particles was not correctly included in Ref.~\cite{hep-ph/0402174}.
Recently, two of the present authors (MK and TT) have applied the technique of thermal field theory to this issue
and found that the effective mass of the order of $H$ for $\phi$ does arise even in RD era~\cite{Kawasaki:2011zi} .

In Ref.~\cite{Kawasaki:2011zi}, the effective mass squared of $\phi$ was expressed in terms of the thermal expectation value of 
the kinetic term of the coupled (scalar  or  fermion) field in the thermal bath.
Here, the thermalized fields were implicitly assumed to be gauge singlets.
Then, the thermal expectation value was evaluated based on thermal field theory.
However, the procedure given in  Ref.~\cite{Kawasaki:2011zi} would become complicated if we considered all the contributions to the effective mass of $\phi$
from a realistic thermal bath consisting of the SUSY SM particles. 
This is because the thermal expectation values of  the gauge covariant kinetic terms would have to be evaluated.

In this paper, we use a different strategy for evaluating the effective mass of $\phi$.
Our observation is as follows: 
the evaluation will become simple and transparent,
if we use the nature of supersymmetry before applying the thermal field theoretic calculation.
Namely, we first rescale  the chiral superfields so that  the $\phi$-dependence 
is absorbed into yukawa and gauge couplings. This enables us to read off 
 the effective mass term for  $\phi$ from the free energy calculated with the rescaled couplings. 

The main purpose of this  paper is to propose a systematic evaluation of the effective mass of $\phi$
and show an example calculation with the minimal supersymmetric standard model (MSSM) plasma.
As a first estimate, we give a complete analytic expression for the leading order (in terms of couplings) contribution to the effective mass of $\phi$. 
The rest of this paper is organized as follows. 
In Sec.~\ref{sec:2}, we explain our strategy for evaluating the effective mass of $\phi$ from yukawa couplings.
In Sec.~\ref{sec:3}, we also explain how to incorporate the contributions of the gauge couplings.
Then, in Sec.~\ref{sec:4}, we give an analytic expression for the effective mass of $\phi$ arising from MSSM plasma.
We also show the numerical result for the temperature dependence of the effective mass of $\phi$.
Sec.~\ref{sec:5} is devoted to conclusions.

\section{Contribution from a yukawa coupling}\label{sec:2}
In this section, we consider a scalar field $\chi$ and a fermion $\tilde \chi$ in the thermal bath in SUSY.
For the moment, we omit the gauge fields for simplicity, though the following procedures can be applied directly to the case with the gauge fields.
The scalar field $\phi$, which is decoupled from the thermal bath, is assumed to have a coupling with $\chi$ and $\tilde \chi$ only through the following non-minimal K$\ddot {\text{a}}$hler potential:
\begin{equation}\label{eq:Kahler}
\begin{split}
K = |\phi|^2 + |\chi|^2 + c \frac{|\phi|^2 |\chi|^2}{M_{\text{P}}^2},
\end{split}
\end{equation}
where $\phi$ and $\chi$ are chiral superfields which include the scalar $\phi$ and the scalar $\chi$, fermion $\tilde \chi$ as component fields, respectively\footnote{
Here and hereafter,  we use the same symbols $\phi$ and $\chi$ for both the superfields and the component scalar fields,
unless otherwise stated. 
}. Here, $M_{\text{P}} \simeq 2.4 \times 10^{18}~\mathrm{GeV}$ is the reduced Planck mass and $c$ is a model dependent parameter,
and we will consider $c =\mathcal{O} (1)$.
In the following subsections, we evaluate the contributions to the effective mass of the scalar field $\phi$, $\tilde m_{\phi}$,  from the thermalized fields $\chi$ and $\tilde \chi$. 
In the evaluation, we neglect the zero-temperature masses of $\chi$ and $\tilde \chi$  for simplicity\footnote{
The following argument is valid when the zero-temperature masses of $\chi$ and $\tilde \chi$ are much smaller than $m_s$ and $m_f$, respectively.
Here, $m_s$ ($m_f$) is the thermal mass of $\chi$ ($\tilde \chi$).
When the zero-temperature masses of $\chi, \tilde \chi$ are comparable to the thermal masses $m_s, m_f$, we have to include contributions from the zero-temperature masses to $\tilde m_{\phi}$.
}.

\subsection{Scalar contributions}
\label{sec:2.1}
In this subsection, we will take into account  the supergravity effects that appear both in the kinetic term of
 the scalar field $\chi$ and in the  F-term potential. We note that the latter effects  were neglected in Ref.~\cite{Kawasaki:2011zi}.

From Eq.~(\ref{eq:Kahler}), the scalar field $\chi$ has the following kinetic term:
\begin{equation}\label{eq:sckinsugra}
\begin{split}
\mathcal{L}_{\text{kin.}}^{\chi} 
&= \left( 1 + \frac{c |\phi|^2}{M_{\text{P}}^2} \right) \partial_{\mu} \chi^* \partial^{\mu} \chi.
\end{split}
\end{equation}
On the other hand, the F-term potential in supergravity is given by the following formula~\cite{350988}:
\begin{equation}\label{eq:scalarp}
\begin{split}
V_F = \mathrm{e}^{K/M_{\text{P}}^2} \left( D_i W K^{i \bar j} \overline{D_j W} - \frac{3 |W|^2}{M_{\text{P}}^2} \right),
\end{split}
\end{equation}
where $W$ is the superpotential, $D_i W = W_i + K_i W / M_{\text{P}}^2$ and $K^{i \bar j}$ is the inverse of $K_{i \bar j}$.
Here the subscript $i$ of $W_i, K_i$ represents the derivative by a scalar field $i$.
Assuming that the superpotential is independent of $\phi$, i.e., $W = W (\chi)$,
we can extract the $\phi$-dependent term from Eq.~(\ref{eq:scalarp}) as
\begin{equation}\label{eq:Vshat}
\begin{split}
V_F |_{\phi \text{-dep.}} = \left( 1 + \frac{(1 - c) |\phi|^2}{M_{\text{P}}^2} \right) |W_{\chi}|^2 + \mathcal{O} \left( M_{\text{P}}^{-4} \right).
\end{split}
\end{equation}
Below, in order to regard the scalar field $\phi$ as a quasi-static external field for $\chi$ (and $\tilde \chi$), 
we assume that the zero-temperature mass of $\phi$ is much smaller than the thermalization rate of $\chi$ (and $\tilde \chi$).
Then, we have the canonical kinetic term for the scalar field $\chi$ by rescaling:
\begin{equation}\label{eq:ress}
\begin{split}
\mathcal{L}_{\text{kin.}}^{\chi} = \partial_{\mu} \hat \chi^* \partial^{\mu} \hat \chi,~~
\hat \chi \equiv \left( 1 + \frac{c |\phi|^2}{M_{\text{P}}^2} \right)^{1/2} \chi.
\end{split}
\end{equation}

Now, we consider the following yukawa interaction in the superpotenital for $\chi$:
\begin{equation}\label{eq:supyu}
\begin{split}
W = \frac{y}{3!} \chi \chi \chi.
\end{split}
\end{equation}
Then, from Eqs.~(\ref{eq:Vshat}) and (\ref{eq:ress}), we obtain
\begin{equation}\label{eq:4yukawa}
\begin{split}
V_F |_{\phi \text{-dep.}}
&= \left( 1 + \frac{ (1 - c) |\phi|^2}{M_{\text{P}}^2} \right) \left( 1 + \frac{c |\phi|^2}{M_{\text{P}}^2} \right)^{-2} \frac{y^2}{4} \left( |\hat \chi|^2 \right)^2
= \frac{\hat y^2}{4} \left( |\hat \chi|^2 \right)^2,
\end{split}
\end{equation}
where we have replaced the coupling $y^2$ with $\hat y^2$ defined by
\begin{equation}\label{eq:resys}
\begin{split}
\hat y^2
\equiv y^2 \left( 1 + \frac{(1 - 3 c) |\phi|^2}{M_{\text{P}}^2} \right).
\end{split}
\end{equation}
Here and hereafter, we neglect $\mathcal{O} (M_{\text{P}}^{-4})$ terms.
Note that, using the canonically normalized scalar field $\hat \chi$, the supergravity effects in the kinetic term~(\ref{eq:sckinsugra}) and the F-term potential~(\ref{eq:Vshat})
are eventually absorbed into the rescaled yukawa coupling $\hat y^2$.
What we have to do for evaluating $\tilde m_{\phi}$ is, then, to extract the effective mass term for $\phi$ from the free energy generated by the rescaled yukawa coupling $\hat y^2$.
The 4-point interaction~(\ref{eq:4yukawa}) gives rise to the 2-loop contribution to the free energy of the system, $\Omega_2$, given by\footnote{
For evaluation of free energy in thermal field theory, for example see Ref.~\cite{Kapsta:2006}. 
}
\begin{equation}
\begin{split}
\Omega_2
= \frac{\hat y^2 T^4}{288}
&= \frac{y^2 T^4}{288} - \frac{\left( c - \frac{1}{3} \right) y^2 |\phi|^2}{96} \frac{T^4}{M_{\text{P}}^2}.
\end{split}
\end{equation}
Thus, we obtain the following effective mass-squared $\tilde m_{\phi}^2$ from the 4-point yukawa interaction~(\ref{eq:4yukawa}):
\begin{equation}\label{eq:ressy}
\begin{split}
\tilde m_{\phi}^2
&= - \frac{c - \frac{1}{3}}{96} \frac{y^2 T^4}{M_{\text{P}}^2}.
\end{split}
\end{equation}
Note that $\tilde m_{\phi}^2$ naturally vanishes for $c=1/3$ in Eq.~(\ref{eq:ressy}),
since $c=1/3$ corresponds to the sequestered K$\ddot {\text{a}}$hler potential form
with which the Planck-suppressed interaction between $\phi$ and $\chi$ is essentially absent.
As a remark of this subsection, when we consider contributions to $\tilde m_{\phi}$ from MSSM plasma in Sec.~\ref{sec:4},
the above procedure is applied to 4-point interactions which consist of squarks, sleptons and Higgs fields.

\subsection{Fermion contributions}
\label{sec:2.2}
In this subsection, we will take into account  the supergravity effects in a yukawa interaction involving the fermion $\tilde \chi$ as well as in the kinetic terms for $\chi$ and $\tilde \chi$.
We note that the supergravity effects in the yukawa interaction were neglected in Ref.~\cite{Kawasaki:2011zi}.

From Eq.~(\ref{eq:Kahler}), the fermionic field $\tilde \chi$ has the following kinetic term:
\begin{equation}
\begin{split}
\mathcal{L}_{\text{kin.}}^{\tilde \chi} 
&= \left( 1 + \frac{c |\phi|^2}{M_{\text{P}}^2} \right) \tilde \chi i \sigma^{\mu} \partial_{\mu} \tilde \chi^*.
\end{split}
\end{equation}
On the other hand, the fermion interaction term in supergravity is given by the following formula~\cite{350988}:
\begin{equation}\label{eq:fsugraint}
\begin{split}
\mathcal{L}_{f}
&= - \frac{1}{2} \mathrm{e}^{K/(2 M_{\text{P}}^2)} \left( \mathcal{D}_i D_j W \right) \xi^i \xi^j + h.c. + \cdots,
\end{split}
\end{equation}
where $\xi^i$ are two-component fermionic fields and $\cdots$ includes interactions between $\xi^i$ and gauge, gravity superfields.
Here, $\mathcal{D}_i D_j W = W_{i j} + K_{i j} W / M_{\text{P}}^2 + K_i D_j W / M_{\text{P}}^2
+ K_j D_i W / M_{\text{P}}^2 - K_i K_j W / M_{\text{P}}^4 - \Gamma^k_{i j} D_k W / M_{\text{P}}^4$
and $\Gamma^k_{i j} = K^{i \bar l} (K_{j \bar l})_i$.
Since $\phi$ is treated as a quasi-static external field for the fermion $\tilde \chi$, we have the canonical kinetic term for $\tilde \chi$ by rescaling:
\begin{equation}\label{eq:resf}
\begin{split}
\mathcal{L}_{\text{kin.}}^{\tilde \chi} 
&= \hat{\tilde \chi} i \sigma^{\mu} \partial_{\mu} \hat{\tilde \chi}^*,
~~\hat{\tilde \chi} = \left( 1 + \frac{c |\phi|^2}{M_{\text{P}}^2} \right)^{1/2} \tilde \chi.
\end{split}
\end{equation}
The rescaling factor coincides with the scalar field case (see Eq.~(\ref{eq:ress})),
since it is actually possible to rescale the superfield $\chi$ to absorb the $\phi$-dependence. 

Assuming the non-minimal K$\ddot {\text{a}}$hler potential~(\ref{eq:Kahler}) and the superpotential~(\ref{eq:supyu}),
Eq.~(\ref{eq:fsugraint}) gives rise to the following $\phi$ - $\tilde \chi$ interaction:
\begin{equation}\label{eq:fsugrayukawa}
\begin{split}
\mathcal{L}_f
&= - \left( 1 + \frac{|\phi|^2}{2 M_{\text{P}}^2} \right) \left( 1 + \frac{c |\phi|^2}{M_{\text{P}}^2} \right)^{-3/2}~\frac{y}{2} \hat \chi \hat{\tilde \chi} \hat{\tilde \chi} + h.c.
= - \frac{\hat y}{2} \hat \chi \hat{\tilde \chi}  \hat{\tilde \chi} + h.c.,
\end{split}
\end{equation}
where $\hat y$ is identical to the one given in Eq.~(\ref{eq:resys})\footnote{
The superficial gap of order $\mathcal{O} \left( M_{\text{P}}^{-4} \right)$ is due to the approximation employed here.
}.
Note that the supergravity effects in the kinetic terms~(\ref{eq:ress}), (\ref{eq:resf}) and
the scalar-fermion-fermion interaction~(\ref{eq:fsugrayukawa}) are absorbed into the rescaled yukawa coupling $\hat y$.
Then, all we need to do is to extract the effective mass term for $\phi$ from the free energy arising from the rescaled yukawa coupling.
The scalar-fermion-fermion interaction~(\ref{eq:fsugrayukawa}) generates the 2-loop contribution to the free energy of the system, $\Omega_2$, given by
\begin{equation}
\begin{split}
\Omega_2
= \frac{5 \hat y^2 T^4}{1152}
= \frac{5 y^2 T^4}{1152} - \frac{(c - \frac{1}{3}) 5 y^2 |\phi|^2}{384} \frac{T^4}{M_{\text{P}}^2}.
\end{split}
\end{equation}
Thus, we obtain the following effective mass-squared $\tilde m_{\phi}^2$ from the yukawa interaction~(\ref{eq:fsugrayukawa}):
\begin{equation}
\begin{split}
\tilde m_{\phi}^2 = - \frac{5 (c - \frac{1}{3})}{384} \frac{y^2 T^4}{M_{\text{P}}^2}.
\end{split}
\end{equation}
Before closing this subsection, we note that when we consider contributions to $\tilde m_{\phi}$ from MSSM plasma in Sec.~\ref{sec:4},
the above procedure is applied to the quark-(s)quark-Higgs(ino) and lepton-(s)lepton-Higgs(ino) yukawa interactions originated from the MSSM superpotential.

\section{Contribution from a gauge coupling}\label{sec:3}
We have seen in Sec.~\ref{sec:2} that  the $\phi$-dependences  are absorbed into 
 the yukawa couplings by the rescaling~(\ref{eq:ress}) and (\ref{eq:resf}). 
In this section, we will see that, if the coupled field is charged under gauge symmetry,
 the rescaling of the chiral fermion generates $\phi$-dependent corrections in the
gauge coupling at one-loop level. As we shall see later,  the numerical coefficient of this correction
turns out to be relatively large especially for the $SU(3)_c$, which partially cancels the one-loop suppression.

In this section, we assume that there are chiral supermultiplets $\chi_i$ which have gauge charges,
and that the corresponding gauge supermultiplet $V = V^a T^a$ ($T^a$ is the generator) in the thermal bath 
interacts with $\phi$ only through the K$\ddot {\text{a}}$hler potential.\footnote{
In particular, no dilatonic coupling is assumed. 
}
The non-minimal K$\ddot {\text{a}}$hler potential~(\ref{eq:Kahler}) is now modified to
\begin{equation}\label{eq:Kahlerg}
\begin{split}
K 
&= |\phi|^2 + \sum_i \left( 1 + \frac{c_i |\phi|^2}{M_{\text{P}}^2} \right) \chi_i^{\dagger} \mathrm{e}^{2 g V} \chi_i,
\end{split}
\end{equation}
where the sum  runs over all the chiral supermultiplets $\chi_i$.
In order to obtain the canonical kinetic term, we rescale the chiral supermultiplets $\chi_i$ as
\begin{equation}\label{eq:resm}
\begin{split}
\hat \chi_i \equiv \left( 1 + \frac{c_i |\phi|^2}{M_{\text{P}}^2} \right)^{1/2} \chi_i.
\end{split}
\end{equation}
Since the chiral supermultiplets $\chi_i$ have the gauge charge, the rescalings~(\ref{eq:resm}) give rise to the following rescaling anomaly~\cite{Konishi:1985tu, ArkaniHamed:1997mj}: 
\begin{equation}
\begin{split}
\prod_i \mathcal{D} \chi_i \mathcal{D} \chi_i^{\dagger}
= \prod_i \mathcal{D} \hat \chi_i \mathcal{D} \hat \chi_i^{\dagger}~\mathrm{exp} \left\{ i \int \mathrm{d}^4 x~\sum_i \frac{- 1}{16} \int \mathrm{d}^2 \theta~\frac{ t_2(\chi_i)}{8 \pi^2} \frac{c_i |\phi|^2}{M_{\text{P}}^2} W_{\alpha}^a (V_h) W^{\alpha a} (V_h) + h.c. \right\},
\end{split}
\end{equation}
where  $t_2 (\chi_i)$ is the Dynkin index and is equal to $1/2$ when $\chi_i$ belongs to the fundamental representation,
and $V_h$ is the gauge supermultiplet with holomorphic gauge coupling.
($V$ and $g$ are the canonically normalized gauge supermultiplet and coupling before the rescaling.)
Here and hereafter, we neglect the $\mathcal{O} (M_{\text{P}}^{-4})$ terms.
Then, the gauge supermultiplet has the following kinetic term:
\begin{equation}
\begin{split}
\mathcal{L}_{\text{kin.}}^{\text{gauge}}
&= \frac{1}{16} \int \mathrm{d}^2 \theta~ \frac{1}{\hat g^2} W_{\alpha}^a (\hat g V) W^{\alpha a} (\hat g V) + h.c.,
\end{split}
\end{equation}
where we have defined the rescaled gauge coupling $\hat g^2$ as
\begin{equation}\label{eq:resg}
\begin{split}
\hat g^2 
= g^2 \left(1 - g^2 \sum_i \frac{t_2 (\chi_i)}{8 \pi^2} \frac{c_i |\phi|^2}{M_{\text{P}}^2} \right)^{-1}
\simeq  g^2 \left(1 + \sum_i \frac{t_2 (\chi_i)}{2 \pi} \frac{g^2}{4 \pi} \frac{c_i |\phi|^2}{M_{\text{P}}^2} \right).
\end{split}
\end{equation}
From Eq.~(\ref{eq:resg}), we see that the rescaled gauge coupling $\hat g^2$ has the $\phi$ dependence but with an extra loop-suppression factor compared to the yukawa coupling contribution~(\ref{eq:resys}).
When $g$ is the gauge coupling constant of an $SU(N_c)$ SUSY Yang-Mills theory, 
the $SU(N_c)$ gauge interactions give rise to the 2-loop contribution to the free energy of the system, $\Omega_2$, which is give by~\cite{Grundberg:1995cu}
\begin{equation}\label{eq:gauge-contri}
\begin{split}
\Omega_2
&= N_g \left( N_c + 3 \sum_i t_2 (\chi_i) \right) \frac{\hat g^2 T^4}{64} \\
&= \left( N_c^2 - 1 \right) \left( N_c + 3 \sum_i t_2 (\chi_i) \right) \left(1 + \sum_i \frac{t_2 (\chi_i)}{2 \pi} \frac{g^2}{4 \pi} \frac{c_i |\phi|^2}{M_{\text{P}}^2} \right) \frac{g^2 T^4}{64},
\end{split}
\end{equation}
where we have used $N_g = N_c^2 - 1$.
Thus, we obtain the following effective mass-squared $\tilde m_{\phi}^2$ generated by the gauge coupling $g$:
\begin{equation}\label{eq:gauge-contri-m}
\begin{split}
\tilde m_{\phi}^2 = \left( N_c^2 - 1 \right) \left( N_c + 3 \sum_i t_2 (\chi_i) \right) \frac{\sum_i c_i t_2 (\chi_i) }{128 \pi} \frac{g^2}{4 \pi} \frac{g^2 T^4}{M_{\text{P}}^2}.
\end{split}
\end{equation}
Note that the numerical coefficient, $(N_c^2 - 1) (N_c + 3 \sum_i t_2 (\chi_i))$, can be large, partially canceling the
the one-loop suppression factor. Therefore we cannot simply neglect the contribution to $\tilde m_{\phi}^2$ 
from the gauge coupling.
In the next section, we evaluate all the 2-loop free energy generated by the rescaled yukawa and gauge couplings in MSSM.
There, we will see that the gauge coupling contributions to $\tilde m_{\phi}$ can be large corrections to the top yukawa coupling contribution.

\section{Hubble-induced mass from MSSM plasma}\label{sec:4}
In this section, we provide an analytic expression for the effective mass $\tilde m_{\phi}$ from the yukawa and gauge couplings in MSSM.
We also estimate the temperature dependence of $\tilde m_{\phi}^2 / H^2$ numerically.

We have explained in the previous sections how to evaluate $\tilde m_{\phi}$ for the given non-minimal K$\ddot {\text{a}}$hler potential and superpotential.
In the following, we assume the non-minimal K$\ddot {\text{a}}$hler potential~(\ref{eq:Kahlerg})
where $i$ is now regarded as the MSSM chiral supermultiplet and we replace $g V$ with the MSSM gauge superfields (times gauge couplings).
We also assume sufficiently high temperature of the plasma and neglect all the zero-temperature (soft SUSY-breaking) masses of MSSM particles\footnote{
We neglect the soft SUSY-breaking masses in the analytic expression for $\tilde m_{\phi}$,
while we take it into account in the renormalization group running of the couplings.
}.

First, let us evaluate the contribution to $\tilde m_{\phi}$ from the MSSM yukawa couplings.
We consider the following MSSM superpotential:
\begin{equation}\label{eq:supp}
\begin{split}
W_{\text{MSSM}}
&= y_t \left( \bar t_R t_L H_u^0 - \bar t_R b_L H_u^+ \right) 
+ y_b \left( \bar b_R b_L H_d^0 - \bar b_R t_L H_d^- \right)
+ y_{\tau} \left( \bar \tau_R \tau_L H_d^0 - \bar \tau_R \nu_{\tau} H_d^- \right),
\end{split}
\end{equation}
where $t_L, b_L, \tau_L, \nu_{\tau}, H_u^+, H_u^0, H_d^0$ and $H_d^-$ are the $SU(2)_L$ charged chiral superfields,
and $\bar t_R, \bar b_R$ and $\bar \tau_R$ are the $SU(2)_L$ singlet anti-particle chiral superfields.
Here, we have omitted the 1st and 2nd generation yukawa couplings since they are are much smaller than the 3rd generation ones.
Now, we include the supergravity effect which we have discussed in section~\ref{sec:2}.
Namely, we rescale all the chiral supermultiplets  and yukawa couplings in order to absorb the supergravity effects in the kinetic terms, F-term potential and fermion interactions into the yukawa couplings $y_t, y_b$ and $y_{\tau}$.
As a consequence, we find that the rescaling results in the following replacement for the yukawa couplings $|y|^2 \to |\hat y|^2$ ($y= y_t, y_b, y_{\tau}$): 
\begin{equation}\label{eq:ythat}
\begin{split}
|\hat y|^2
= |y|^2 \left( 1 + \frac{(1 - c_i - c_j - c_k) |\phi|^2}{M_{\text{P}}^2} \right),
\end{split}
\end{equation}
where $c_i, c_j$ and $c_k$ are the coefficients in the non-minimal K$\ddot {\text{a}}$hler potential~(\ref{eq:Kahler}) for the corresponding chiral fields.
As an illustration, let us consider the interactions arising from the term $W = y_{\tau} \bar \tau_R \tau_L H_d^0$ in Eq.~(\ref{eq:supp}).
From this term, we obtain a 4-point yukawa interaction $|\partial W / \partial \tau_L|^2 = y_{\tau}^2 |\tilde{\bar \tau}_R|^2 |H_d^0|^2$ ($\tilde{\bar \tau}_R, H_d^0$ are the scalar component of the superfields $\bar \tau_R, H_d^0$).
In this case, the coefficients in Eq.~(\ref{eq:ythat}) are determined as $c_i = c_{\tau_L},~c_j = c_{\bar \tau_R},~c_k = c_{H_d^0}$.
On the other hand, one of the yukawa interactions involving fermions we obtain from the term $W$ is $- y_{\tau} \bar \tau_R \tau_L H_d^0 + h.c.$ ($\bar \tau_R, \tau_L$ are fermions and $H_d^0$ is a scalar).
For this interaction, we determine the coefficients in Eq.~(\ref{eq:ythat}) as $c_i = c_{\tau_L},~c_j = c_{\bar \tau_R},~c_k = c_{H_d^0}$ which are identical to the above 4-point yukawa interaction contribution.
Now, taking into account of Eq.~(\ref{eq:ythat}), the sum of the 2-loop contributions to the free energy, $\Omega_2$, from the 3rd generation yukawa couplings are summarized as following: 
\begin{equation}\label{eq:yukawa}
\begin{split}
\Omega_2|_{\text{yukawa}}
&= \frac{9 \pi T^4}{8} \sum_{i = t, b, \tau} \gamma_i \alpha_{y_i} \left( 1 - \frac{3 (\bar c_i - \frac{1}{3}) |\phi|^2}{M_{\text{P}}^2} \right).
\end{split}
\end{equation}
where we have defined $\alpha_{y_i} \equiv |y_i|^2/(4\pi)$, $\gamma_t = \gamma_b = 1$ and $\gamma_{\tau} = 1/3$.
Here, $\bar c_t, \bar c_b$ and $\bar c_{\tau}$ are defined by $\bar c_t = \frac{1}{3} \left( c_{\bar t_R} + c_{t_L} + c_{H_u} \right)$,
$\bar c_b = \frac{1}{3} \left( c_{\bar b_R} + c_{t_L} + c_{H_d} \right)$
and $\bar c_{\tau} = \frac{1}{3} \left( c_{\bar \tau_R} + c_{\tau_L} + c_{H_d} \right)$, respectively.
Since the chiral superfields which are included in the same gauge multiplet should have the same coefficient $c_i$,
we have set $c_{b_L} = c_{t_L}, c_{H_u^0} = c_{H_u^+}, c_{H_d^0} = c_{H_d^-}, c_{\nu_{\tau}} = c_{\tau_L}$.
Each contribution to $\Omega_2$ is briefly described in Appendix.
From Eq.~(\ref{eq:yukawa}), we can extract the contribution to $\tilde m_{\phi}^2$ from the yukawa couplings $y_t, y_b$ and $y_{\tau}$.

Next, let us evaluate the contribution to $\tilde m_{\phi}$ from the MSSM gauge couplings.
Using the formula Eq.~(\ref{eq:resg}), we obtain the rescaled gauge couplings in MSSM as
\begin{equation}\label{eq:rescg}
\begin{split}
&\hat \alpha_s = \alpha_s \left ( 1 + \frac{3}{\pi} \frac{\bar c_s \alpha_s |\phi|^2}{M_{\text{P}}^2} \right),
~\hat \alpha_2 = \alpha_2 \left( 1 + \frac{7}{2 \pi} \frac{\bar c_2 \alpha_2 |\phi|^2}{M_{\text{P}}^2} \right),
~\hat \alpha_Y = \alpha_Y \left( 1 + \frac{11}{2 \pi} \frac{\bar c_Y \alpha_Y |\phi|^2}{M_{\text{P}}^2} \right).
\end{split}
\end{equation}
And, from Eq.~(\ref{eq:gauge-contri-m}), the resultant contribution to $\tilde m_{\phi}^2|_{2\text{-loop}}$ is summarized as follows
\begin{equation}\label{eq:su}
\begin{split}
&\tilde m_{\phi}^2|_{SU(3)_c} = \frac{63}{2} \frac{\bar c_s \alpha_s^2 T^4}{M_{\text{P}}^2},
~~\tilde m_{\phi}^2|_{SU(2)_L} = \frac{483}{32} \frac{\bar c_2 \alpha_2^2 T^4}{M_{\text{P}}^2},
~~\tilde m_{\phi}^2|_{U(1)_Y} = \frac{363}{32} \frac{\bar c_Y \alpha_Y^2 T^4}{M_{\text{P}}^2},
\end{split}
\end{equation}
where $\alpha_i = g_i^2/(4 \pi)$, and $g_s, g_2$ and $g_Y$ are the gauge couplings of $SU(3)_c, SU(2)_L$ and $U (1)_Y$, respectively.
Here, we have defined $\bar c_s = \frac{1}{12} \sum_{i}^{SU(3)_c \text{triplet}} c_i$,
$\bar c_2 = \frac{1}{14} \sum_{i}^{SU(2)_L \text{doublet}} c_i$ and
$\bar c_Y = \sum_i Y_i^2 c_i / \sum_i Y_i^2$.
In the definition of $\bar c_Y$, $i$ runs all the $U(1)_Y$ chiral supermultiplets.

Now, we are in a position to evaluate the total amount of the effective mass of the Planck-suppressed interacting scalar field $\phi$.
From Eqs.~(\ref{eq:yukawa}) and (\ref{eq:su}),  the total contribution to $\tilde m_{\phi}^2$ from the MSSM plasma is given by
\begin{equation}\label{eq:sum}
\begin{split}
\tilde m_{\phi}^2
&= - \frac{27 \pi}{8} \sum_{i=t,b,{\tau}} \gamma_i \left( \bar c_i - \frac{1}{3} \right) \alpha_{y_i} \frac{T^4}{M_{\text{P}}^2}
+ \left( \frac{63}{2} \bar c_s \alpha_s^2 + \frac{483}{32} \bar c_2 \alpha_2^2 + \frac{363}{32} \bar c_Y \alpha_Y^2  \right) \frac{T^4}{M_{\text{P}}^2} \\
&= \left\{ - \frac{81}{61 \pi} \sum_{i=t,b,{\tau}} \gamma_i \left( \bar c_i - \frac{1}{3} \right) \alpha_{y_i}
+ \frac{756}{61 \pi^2} \bar c_s \alpha_s^2 + \frac{1449}{244 \pi^2} \bar c_2 \alpha_2^2 + \frac{1089}{244 \pi^2} \bar c_Y  \alpha_Y^2 \right\} H^2,
\end{split}
\end{equation}
where, in the second line, we have used the Friedmann equation in RD era $3 M_{\text{P}}^2 H^2 = \frac{\pi^2 g_*}{30} T^4$
and $g_* = 228.75 = 915/4$ for the MSSM plasma as the relativistic degrees of freedom in the thermal bath.
Here, the thermalization rate of the MSSM particles are much larger than the Hubble expansion rate,
and thus the Hubble expansion rate is involved in the above evaluation only through the temperature of the thermal bath.
Furthermore, assuming the thermal bath is large enough, we neglect the backreaction of $\phi$'s interaction to the thermal bath.
Eq.~(\ref{eq:sum}) is the analytic expression for $\tilde m_{\phi}^2$ and is the main result of this paper.
Note that the largest contributions to $\tilde m_{\phi}^2$ come from the top yukawa coupling $y_t$ and $SU(3)_c$ gauge coupling $g_s$ in typical temperature (see figures below).

Let us comment on  higher-loop contributions to $\tilde m_{\phi}^2$.
In QCD at finite temperature, it has been recognized that the higher-loop contributions to the free energy are important and the convergence is poor in the ordinary perturbation theory.
A lot of effort has been paid to the calculation of the higher-loop contributions to the free energy and even the $\mathcal{O} (g_s^6 \ln g_s)$ result was  obtained~\cite{Kajantie:2002wa}.
On the other hand, improved perturbation theories are also investigated
and the resultant free energies are found to have good convergence~\cite{Andersen:1999fw} (for reviews see Ref.~\cite{Blaizot:2003tw}).
From the results in these studies, we observe that the leading order result in the ordinary perturbation theory is different from the convergence-improved result at most by a factor of order unity. 
Returning to our subject, the poor convergence of the free energy evaluation in the ordinary perturbation theory would be true also in MSSM at finite temperature.
In fact, we evaluate the next-to-leading order contributions to $\tilde m^2_{\phi}$ in Appendix B, and find that the next-to-leading order contribution is comparable to the leading order (2-loop) one.
However, from the observation in the QCD results, even if we include the higher-loop contributions,
the magnitude of $\tilde m_{\phi}^2$ would change from the leading order one at most by a factor of order unity.
Thus, the leading order result~(\ref{eq:sum}) can serve as the first estimate of $\tilde m^2_{\phi}$ from the MSSM plasma.
Since our main purpose of this paper is to propose a systematic evaluation of $\tilde m_{\phi}$ and show an example calculation with the MSSM plasma, we do not pursue the effect of the higher-loop contributions on $\tilde m_{\phi}$ here.

Lastly we show the numerical results for the temperature dependence of $\tilde m_{\phi}^2/H^2$. 
In all the figures, we use the public code SOFTSUSY~\cite{Allanach:2001kg} in order to evolve the coupling constants according to the renormalization group equations.
For the sake of simplicity, we apply the boundary condition of the minimal supergravity model.
However, it should be emphasized that the resultant value of $\tilde m_{\phi}$ does not change significantly
even if we impose other boundary condition like the minimal gauge-mediated SUSY breaking or minimal anomaly-mediated SUSY breaking model one.
Below,  we take $m_0=m_{1/2}=3~\mathrm{TeV},~A_0 = 0,~\tan \beta = 20~(\text{and}~40),~\mathrm{sign} (\mu) = +1$ in the minimal supergravity model.
Here, $m_0, m_{1/2}$ and $A_0$ are the unified scalar mass, gaugino mass, trilinear scalar coupling at the GUT scale, respectively,
and $\mathrm{sign} (\mu)$ is the sign of the supersymmetric $\mu$ term. 
Also, $\tan \beta = \langle H_u^0 \rangle / \langle H_d^0 \rangle$ is the ratio of the Higgs field vacuum expectation values at the weak scale.
$\tilde m_{\phi}^2 / H^2$ has only small dependence on the parameter choice as long as the soft SUSY-breaking masses are $\mathcal{O} (1 \sim 10)~\mathrm{TeV}$.

\begin{figure}[t]
\begin{center}
\includegraphics[width=75mm]{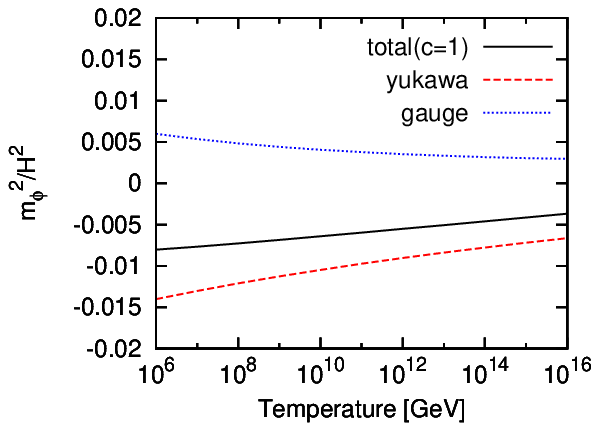}
\includegraphics[width=75mm]{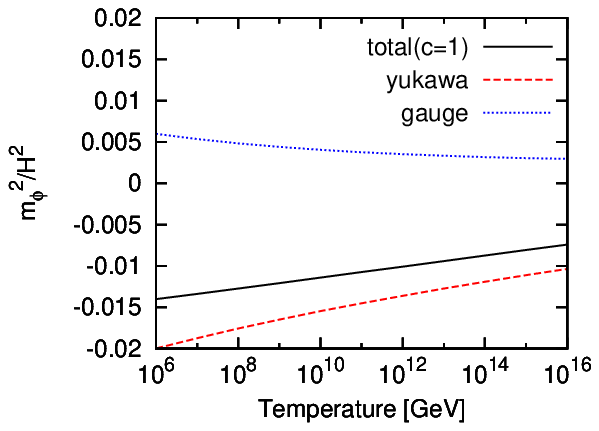}
\end{center}
\caption{Hubble-induced mass-squared $\tilde m_{\phi}^2$ for $\tan \beta = 20~(\text{left panel}),~40~(\text{right panel})$.
We set $c_i=1$ for all chiral superfields $i$.
The black solid line is the total (yukawa + gauge) contribution to $\tilde m_{\phi}^2/H^2$.
The red dashed line, blue dotted line are the sum of the yukawa, gauge coupling contributions to $\tilde m_{\phi}^2/H^2$, respectively.
}
\label{fig:c1}
\end{figure}
Fig.~\ref{fig:c1} shows $\tilde m_{\phi}^2/H^2$ for $\tan \beta = 20, 40$.
Here, we set $c_i=1$ for all the chiral superfields $i$.
The black solid line is the total (yukawa + gauge) contributions to $\tilde m_{\phi}^2/H^2$.
The red dashed line, blue dotted line are the sum of the yukawa, gauge coupling contributions to $\tilde m_{\phi}^2/H^2$, respectively.
From Fig.~\ref{fig:c1}, one can see that $|\tilde m_{\phi}^2|$ is about $H^2/100$,
though $\tilde m_{\phi}^2$ mildly depends on the plasma temperature and $\tan \beta$.
Although we do not show here, we have checked that the largest contributions to $\tilde m_{\phi}^2$ come from $y_t$ and $g_s$ in most of the temperature range.

\begin{figure}[t]
\begin{center}
\includegraphics[width=75mm]{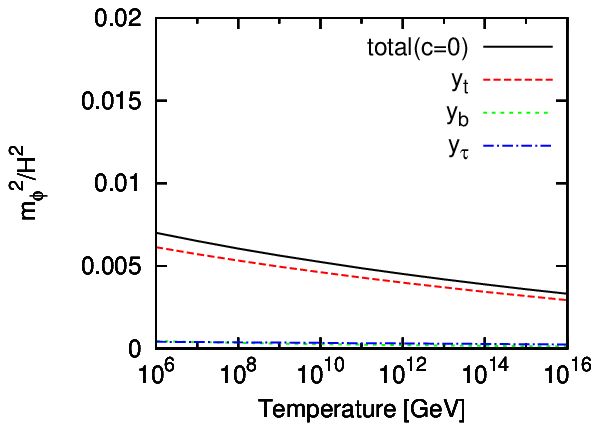}
\includegraphics[width=75mm]{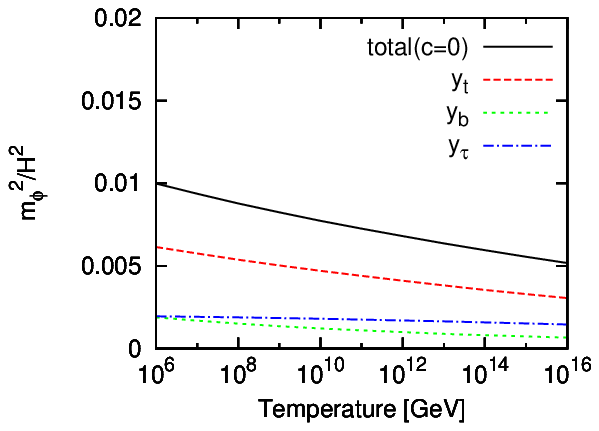}
\end{center}
\caption{Hubble-induced mass-squared $\tilde m_{\phi}^2$ for $\tan \beta = 20~(\text{left panel}),$ $~40~(\text{right panel})$.
We set $c_i=0$ (minimal K$\ddot{\text{a}}$hler potential case) for all chiral superfields $i$.
Here, all the gauge coupling contributions are absent because of the choice $c_i = 0$.
The black solid line, red dashed line,  green dotted line and blue dash-dotted line are the total,
$y_t, y_b$ and $y_{\tau}$ contributions to $\tilde m_{\phi}^2/H^2$, respectively.
}
\label{fig:c0}
\end{figure}
In Fig.~\ref{fig:c0}, we set  $c_i=0$ (minimal K$\ddot{\text{a}}$hler potential case) for all the chiral superfields $i$.
Here, we again choose $\tan \beta = 20, 40$ cases.
The black solid line, red dashed line,  green dotted line and blue dash-dotted line are the total,
$y_t, y_b$ and $y_{\tau}$ contributions to $\tilde m_{\phi}^2/H^2$, respectively. Note that the gauge coupling
contributions vanish since no rescaling of the coupled fields is required. 
From Figs.~\ref{fig:c0}, one can see that $\tilde m_{\phi}^2$ is about $H^2/100$.
We note that $\tilde m_{\phi}^2$ is always positive in this minimal K$\ddot{\text{a}}$hler potential case ($c_i =0$).

\begin{figure}[t]
\begin{center}
\includegraphics[width=80mm]{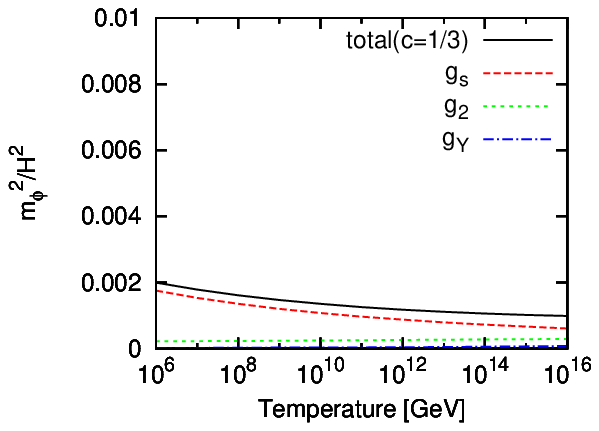}
\caption{Hubble-induced mass-squared $\tilde m_{\phi}^2$.
We have set $c_i=1/3$ (sequestered K$\ddot{\text{a}}$hler potential case) for all chiral superfields $i$.
In this case, all the yukawa coupling contributions vanish and only gauge couplings contribute to $\tilde m_{\phi}^2$.
Here, $\tilde m_{\phi}^2$ has a negligible dependence on $\tan \beta$.
The black solid line, red dashed line,  green dotted line and blue dash-dotted line are the total,
$g_s, g_2$ and $g_Y$ contributions to $\tilde m_{\phi}^2/H^2$, respectively.
}
\label{fig:c03}
\end{center}
\end{figure}
Finally, in Fig.~\ref{fig:c03}, we set  $c_i=1/3$ (sequestered K$\ddot{\text{a}}$hler potential case) for all the chiral superfields $i$.
The black solid line, red dashed line,  green dotted line and blue dash-dotted line are the total,
$g_s, g_2$ and $g_Y$ contributions to $\tilde m_{\phi}^2/H^2$, respectively. The yukawa coupling contributions vanish
in this case since the chiral superfields are essentially decoupled from the scalar $\phi$.  Nevertheless the gauge coupling
contributions appear because of the rescaling anomaly. 
From Fig.~\ref{fig:c03}, one can see that $\tilde m_{\phi}^2$ is about $H^2/1000 \sim H^2/500$.
We note that $\tilde m_{\phi}^2$ is independent of $\tan \beta$ and is always positive in this sequestered K$\ddot{\text{a}}$hler potential case ($c_i =1/3$).

\section{Conclusions}\label{sec:5}
In this paper, we have proposed a systematic evaluation of the effective mass of a Planck-suppressed interacting scalar field $\phi$.
As a concrete and realistic example, we have shown the calculation of $\tilde m_{\phi}$ arising from the MSSM plasma through 
Planck-supressed interactions  in  the non-minimal K$\ddot {\text{a}}$hler potential like~Eq.~(\ref{eq:Kahlerg}).
The strategy we have used is as follows.
First, we have rescaled the chiral superfields so that  the supergravity effects in the kinetic terms, F-term potential and fermion yukawa interactions 
are absorbed into the rescaled yukawa and gauge couplings.
The gauge couplings receive $\phi$-dependent corrections from the rescaling anomaly, which is accompanied by
a one-loop suppression factor (see Eq.~(\ref{eq:resg})) compared to the yukawa couplings~(\ref{eq:ythat}).
However, there are relatively large numerical factors in the rescaled gauge couplings and thus we have to include the gauge coupling contributions in the evaluation of $\tilde m_{\phi}$.
Then invoking the free energy with the rescaled couplings, we have read off the expression for $\tilde m_{\phi}$.
The resultant $\tilde m_{\phi}^2$ arising from the sufficiently high temperature MSSM plasma is given in Eq.~(\ref{eq:sum}), which is about $10^{-3} H^2 \sim 10^{-2} H^2$ for typical parameter sets.
This is our main result in this study.
As we have mentioned below Eq.~(\ref{eq:sum}), the higher-loop contribution to $\tilde m_{\phi}^2$ would be significant.
Thus in order to obtain a reliable result for $\tilde m_{\phi}^2$, we have to proceed the evaluation up to sufficiently higher-loop order or apply the improved perturbation theory.
However, since the leading order result in ordinary perturbation theory would be different from the convergence-improved result at most by a factor, 
the leading order result~(\ref{eq:sum}) can serve as the first estimate of the systematic evaluation of $\tilde m^2_{\phi}$ from the MSSM plasma.
Also, since our main purpose of this paper is to propose the systematic evaluation for $\tilde m_{\phi}$ and show the example calculation with the MSSM plasma,
we don't pursue the higher-loop contribution here.

Before closing this paper, let us briefly discuss the impact of the ``Hubble-induced mass" in RD era~(\ref{eq:sum}) on cosmology.
In superstring and supergravity theories, there are generically light moduli fields, which cause
serious cosmological moduli problem~\cite{Coughlan:1983ci}. One of the attractive solutions is the adiabatic solution~\cite{Linde:1996cx,Nakayama:2011wqa};
if the modulus field receives an enhanced Hubble-induced mass-squared,  $\tilde m_{\phi}^2 = \mathcal{O} (100) H^2$, it follows the time-dependent minimum
and as a result, its abundance is suppressed by a power of the ratio of the zero-temperature modulus mass and the inflation scale~\cite{Nakayama:2011wqa}. 
The origin of such enhanced Hubble-induced mass may be due to a cut-off scale one order of magnitude lower than the Planck scale~\cite{Takahashi:2011as}, or
some strong dynamics at the Planck scale~\cite{Takahashi:2010uw}. Our findings show that, even if the couplings between the modulus and
the MSSM sector are enhanced by two orders of magnitude, i.e., $|c| = {\cal O}(100)$, the Hubble-induced mass for the modulus is not
sufficient to suppress the modulus abundance when it starts to oscillate after reheating. This results in a rather robust upper bound on the
reheating temperature~\cite{Takahashi:2011as, Nakayama:2011wqa} for the adiabatic solution to work.

\section*{Acknowledgments}
TT is grateful to Masahiro Ibe for helpful discussions about the renormalization group method.
This work was supported by the Grant-in-Aid for Scientific research from the Ministry of Education, Science, Sports, and Culture (MEXT), Japan, (No.14102004 and No.21111006) [MK],
Scientific Research on Innovative Areas (No.24111702, No. 21111006, and No.23104008) [FT],
Scientific Research (A) (No. 22244030 and No.21244033) [FT],
and JSPS Grant-in-Aid for Young Scientists (B) (No. 24740135) [FT].
This work was also supported by World Premier International Center Initiative (WPI Program), MEXT, Japan.
The work of TT was supported in part by JSPS Research Fellowships for Young Scientists.

\section*{Appendix A: 2-loop free energy from the yukawa and gauge couplings}
In this appendix, we show each contribution to the 2-loop free energy
from the yukawa couplings~(\ref{eq:yukawa}) and the gauge couplings~(\ref{eq:gauge-contri}).

First, we write down the contributions to Eq.~(\ref{eq:yukawa}):
\begin{equation}
\begin{split}
&\Omega_2^{s s s s} = \frac{|y|^2 T^4}{144} \times 
\begin{cases}
&3~~(\text{with squark}), \\
&1~~(\text{without squark}),
\end{cases} \\
&\Omega_2^{s f f}= \frac{|y|^2 T^4}{144} \times 
\begin{cases}
&\frac{15}{4}~~(\text{with squark}), \\
&\frac{5}{4}~~(\text{without squark}),
\end{cases}
\end{split}
\end{equation}
where, $\Omega_2^{s s s s}$, $\Omega_2^{s f f}$ are the 2-loop contributions to the free energy 
which are generated by the 4-point scalar interaction $s s s s$,  the scalar-fermion-fermion interaction $s f f$, respectively.
Here, we have used $s, f$ as symbols for the scalars and fermions in the relevant interactions. 
There are six 2-loop diagrams from 4-point scalar interactions for each yukawa coupling ($|y_t|^2, |y_b|^2, |y_{\tau}|^2$).
Also, there are six 2-loop diagrams from scalar-fermion-fermion interactions for each yukawa coupling.
After taking the sum of these contributions, we finally obtain Eq.~(\ref{eq:yukawa}).

Next, we write down each 2-loop contribution to Eq.~(\ref{eq:gauge-contri}) from SUSY $SU(N_c)$ theory:
\begin{equation}\label{eq:eachsusync}
\begin{split}
&\Omega_2^{A f f} = N_g \left( \sum_i t_2 (i) \right) \times \frac{5}{4} \times \frac{g^2 T^4}{144},
~~~~\Omega_2^{s f \lambda} = N_g \left( \sum_i t_2 (i) \right) \times \frac{5}{2} \times \frac{g^2 T^4}{144}, \\
&\Omega_2^{s s s s} = N_g \left( \sum_i t_2 (i) \right) \times \frac{1}{2} \times \frac{g^2 T^4}{144},
~~~~\Omega_2^{s s A} = N_g \left( \sum_i t_2 (i) \right) \times \frac{-3}{2} \times \frac{g^2 T^4}{144}, \\
&\Omega_2^{s s A A} = N_g \left( \sum_i t_2 (i) \right) \times 4 \times \frac{g^2 T^4}{144}, \\
&\Omega_2^{A \lambda \lambda} = N_g N_c \times \frac{5}{4} \times \frac{g^2 T^4}{144}, 
~~~~\Omega_2^{A c c} = N_g N_c \times \frac{1}{4} \times \frac{g^2 T^4}{144},\\
&\Omega_2^{AAA} = N_g N_c \times \frac{-9}{4} \times \frac{g^2 T^4}{144},
~~~~\Omega_2^{AAAA} = N_g N_c \times 3 \times \frac{g^2 T^4}{144}.
\end{split}
\end{equation}
where, $\Omega_2^{\hat{\mathcal{O}}}$ is  the 2-loop contribution to the free energy which is generated by the interaction $\hat{\mathcal{O}}$.
Here, we have used $s, f, A, \lambda$ and $c$ as symbols for the chiral scalar, chiral fermion, gauge field, gaugino and ghost field in the relevant interactions, respectively. 
Summing up the contributions in Eq.~(\ref{eq:eachsusync}), we eventually obtain Eq.~(\ref{eq:gauge-contri}).
Here, the summation $\sum_i$ runs all the $SU(N_c)$ chiral supermultiplet $i$.
The Dynkin index $t_2 (i) = 1/2$ when the chiral supermultiplet $i$ belongs to the fundamental representation.
We note that for SUSY $U(1)_Y$ theory, we can apply the formula~(\ref{eq:gauge-contri}) with $N_g = 1, N_c = 0, t_2 (i) = Y_i^2$.

\section*{Appendix B: The next-to-leading order contributions}
In this appendix, we derive an analytic expression for the next-to-leading order contribution to $\tilde m_{\phi}$ from the MSSM plasma.
To do this, we evaluate the contribution to the free energy from the ring diagrams~\cite{GellMann:1957zza, Kapsta:2006}
generated by the rescaled couplings~(\ref{eq:ythat}) and (\ref{eq:rescg}).
Then, from the expression for the ring diagram free energy, we read off the effective mass $\tilde m_{\phi}$ at next-to-leading order.

First, we consider the contribution from the ring diagrams of gluon, $W$-boson and $B$-boson in MSSM to the free energy, $\Omega_3^{\text{gauge}}$.
These gauge field ring diagrams can be evaluated by the usual method in thermal field theory~\cite{Kapsta:2006} and is given by~\cite{Grundberg:1995cu}
\begin{equation}\label{eq:gaugeringd}
\begin{split}
\Omega_3^{\text{gauge}}
&= - \frac{T}{12 \pi} \left( 8 \hat m_{D, g}^3 + 3 \hat m_{D, W}^3 + \hat m_{D, B}^3 \right),
\end{split}
\end{equation}
where $\hat m_{D, g}, \hat m_{D, W}$ and $\hat m_{D, B}$ are the Debye masses of gluon, $W$-boson and $B$-boson, respectively, and are given by~\cite{Comelli:1996vm}
\begin{equation}\label{eq:selfgauge}
\begin{split}
&\hat m^2_{D, g}
= \frac{9}{2} \hat g_s^2 T^2,~~
\hat m^2_{D, W}
= \frac{9}{2} \hat g_2^2 T^2,~~
\hat m^2_{D, B}
= \frac{11}{2} \hat g_Y^2 T^2.
\end{split}
\end{equation}
Note that we have already rescaled the chiral superfields.
Now, using Eqs.~(\ref{eq:rescg}) and (\ref{eq:selfgauge}), the gauge field ring diagram contribution~(\ref{eq:gaugeringd}) reduces to
\begin{equation}\label{eq:omega3gauge}
\begin{split}
\Omega_3^{\text{gauge}}
&= - \frac{T^4}{12 \pi} \left\{ 54 \sqrt{2} \hat g_s^3 + \frac{81}{2 \sqrt{2}} \hat g_2^3 + \frac{11 \sqrt{11}}{2 \sqrt{2}} \hat g_Y^3 \right\} \\
&= - \left\{ \frac{162 \sqrt{2}}{\sqrt{\pi}} \bar c_s \alpha_s^{5/2} + \frac{567}{4 \sqrt{2 \pi}} \bar c_2 \alpha_2^{5/2} + \frac{121 \sqrt{11}}{4 \sqrt{2 \pi}} \bar c_Y \alpha_Y^{5/2} \right\} \frac{T^4}{M_{\text{P}}^2} |\phi|^2
+ (\phi\text{-indep.}).
\end{split}
\end{equation}
Thus, we obtain the following contribution to $\tilde m_{\phi}^2$ from the gauge field ring diagrams in MSSM:
\begin{equation}\label{eq:mringgauge}
\begin{split}
\tilde m_{\phi}^2|^{\text{gague}}_{\text{ring}}
&= - \left\{ \frac{162 \sqrt{2}}{\sqrt{\pi}} \bar c_s \alpha_s^{5/2} + \frac{567}{4 \sqrt{2 \pi}} \bar c_2 \alpha_2^{5/2} + \frac{121 \sqrt{11}}{4 \sqrt{2 \pi}} \bar c_Y \alpha_Y^{5/2} \right\} \frac{T^4}{M_{\text{P}}^2} \\
&= - \left\{ \frac{3888 \sqrt{2}}{61 \pi^{5/2}} \bar c_s \alpha_s^{5/2} + \frac{1701 \sqrt{2}}{61 \pi^{5/2}} \bar c_2 \alpha_2^{5/2} + \frac{363 \sqrt{22}}{61 \pi^{5/2}} \bar c_Y \alpha_Y^{5/2} \right\} H^2,
\end{split}
\end{equation}
where we have used the Friedmann equation in the RD era $3 M_{\text{P}}^2 H^2 = \frac{\pi^2 g_*}{30} T^4$ and $g_* = 228.75 = 915/4$ as in Eq.~(\ref{eq:sum}).
We note that the numerical coefficients of the gauge couplings in Eq.~(\ref{eq:mringgauge}) are about five times larger than the ones in Eq.~(\ref{eq:su}) and have opposite sign.

Next, let us evaluate the contribution from the ring diagrams of the MSSM chiral scalar fields to the free energy, $\Omega_3^{\text{scalar}}$.
The usual method in thermal field theory~\cite{Kapsta:2006} can be applied for the evaluation of $\Omega_3^{\text{scalar}}$ and the result is given by~\cite{Grundberg:1995cu}\footnote{
In Ref.~\cite{Grundberg:1995cu}, the factor $\frac{N_g}{12 N_c}$ should be replaced by $\frac{N_g}{4 N_c}$ in Eqs.(5) and (6) .
This corrected factor $\frac{N_g}{4 N_c}$ agrees with Ref.~\cite{Comelli:1996vm}.
}
\begin{equation}\label{eq:ringscalari}
\begin{split}
\Omega_3^{\text{scalar}}
&= - \frac{T}{6 \pi} \sum_i^{\text{scalar}} m^3_i,
\end{split}
\end{equation}
where $i$ runs all the chiral scalar fields in MSSM.
Here, $m_i$ is the thermal mass of the scalar field $i$ and is summarized in Ref.~\cite{Comelli:1996vm}. 
From Eqs.~(\ref{eq:ythat}), (\ref{eq:rescg})  and (\ref{eq:ringscalari}), after rescaling the MSSM chiral superfields,
the ring diagrams of the chiral scalar fields contribute to the free energy as
\begin{equation}\label{eq:ringscalar}
\begin{split}
\Omega_3^{\text{scalar}}
&= - \sum_i^{\text{scalar}} \frac{\xi_i}{4 \pi} \frac{m_i^3 T}{M_{\text{P}}^2} |\phi|^2 + (\phi\text{-indep.}),
\end{split}
\end{equation}
where $\mathcal{O} (M_{\text{P}}^{-4})$ terms are neglected and $\xi_i$ is defined by
\begin{equation}
\begin{split}
\xi_i \frac{|\phi|^2}{M_{\text{P}}^2} = \frac{\hat m^2_i|_{|\phi|^2}}{m^2_i}.
\end{split}
\end{equation}
Here, $\hat m^2_i|_{|\phi|^2}$ is the $\phi$-dependent part of the thermal mass-squared $\hat m^2_i$ in which the couplings are replaced by the rescaled ones~(\ref{eq:ythat}) and (\ref{eq:rescg}).
From Eq.~(\ref{eq:ringscalar}), we obtain the following contribution to $\tilde m_{\phi}^2$ from the scalar field ring diagrams in MSSM: 
\begin{equation}\label{eq:mringscalar}
\begin{split}
\tilde m_{\phi}^2|^{\text{scalar}}_{\text{ring}}
&= - \sum_i^{\text{scalar}} \frac{\xi_i}{4 \pi} \frac{m_i^3 T}{M_{\text{P}}^2} 
= - \sum_i^{\text{scalar}} \frac{6 \xi_i}{61 \pi^3} \frac{m_i^3}{T^3} H^2.
\end{split}
\end{equation}
Here, we have used the Friedmann equation in the RD era as in Eq.~(\ref{eq:sum}).

Now, we are in a position to sum up the ring diagram contributions.
From Eqs.~(\ref{eq:mringgauge}) and (\ref{eq:mringscalar}), the total ring diagram contribution $\tilde m_{\phi}^2|_{\text{ring}}$ is obtained as follows
\begin{equation}\label{eq:resring}
\begin{split}
\tilde m_{\phi}^2|_{\text{ring}}
= \left\{ \sum_i^{\text{scalar}} \frac{- 6 \xi_i}{61 \pi^3} \frac{m_i^3}{T^3}
 - \frac{3888 \sqrt{2}}{61 \pi^{5/2}} \bar c_s \alpha_s^{5/2} - \frac{1701 \sqrt{2}}{61 \pi^{5/2}} \bar c_2 \alpha_2^{5/2} - \frac{363 \sqrt{22}}{61 \pi^{5/2}} \bar c_Y \alpha_Y^{5/2} \right\} H^2.
\end{split}
\end{equation}
We note that the ring diagram contribution~(\ref{eq:resring}) is rather significant compared with the leading order (2-loop) one given in Eq.~(\ref{eq:sum}).
This would be the signature of the poor convergence of the ordinary perturbation theory as mentioned below Eq.~(\ref{eq:sum}).
Thus in order to obtain a reliable result for $\tilde m_{\phi}^2$, we have to proceed the evaluation up to sufficiently higher-loop order or apply the improved perturbation theory.
However, since the leading order result in ordinary perturbation theory would be different from the convergence-improved result at most by a factor of order unity as we observe in literatures like Ref.~\cite{Andersen:1999fw}, 
the leading order result~(\ref{eq:sum}) can serve as the first estimate of the systematic evaluation of $\tilde m^2_{\phi}$ from the MSSM plasma.

{}

\end{document}